\documentclass[apj]{emulateapj}
\usepackage[ngerman,english]{babel}
\newif\ifAMStwofonts
\AMStwofontstrue 
\def\degr{\hbox{$^\circ$}}

\def\arcsec{\hbox{$^{\prime\prime}$}}

\def\al{{a_{lm}}}

\def\gsim{~\rlap{$>$}{\lower 1.0ex\hbox{$\sim$}}}

\def\simpropto{\lower.2ex\hbox{$\; \buildrel \propto \over \sim \;$}}
\def\ltsim{\lower.5ex\hbox{$\; \buildrel < \over \sim \;$}}
\def\gtsim{\lower.5ex\hbox{$\; \buildrel > \over \sim \;$}}
\def\ltsim{\lower.5ex\hbox{$\; \buildrel < \over \sim \;$}}
\def\gtsim{\lower.5ex\hbox{$\; \buildrel > \over \sim \;$}}

\def\sun{\Theta_{_{\rm un}}}

\def\dd{\,{\rm d}}

\def\dd{{\rm d}}

\def\pmb#1{\setbox0=\hbox{#1}%
\kern-.025em\copy0\kern-\wd0
\kern.05em\copy0\kern-\wd0
\kern-.025em\raise.0433em\box0}

\def\vr{\pmb{r}}

\def\hvn{\hat {\vr}}

\def\vk{\pmb{k}}

\def\simlt{\lower.5ex\hbox{$\; \buildrel < \over \sim \;$}}
\def\simgt{\lower.5ex\hbox{$\; \buildrel > \over \sim \;$}}

\newcommand{\beq}{\begin{equation}}

\newcommand{\eeq}{\end{equation}}
\def\beqa{\begin{eqnarray}}
\def\eeqa{\end{eqnarray}}
\def\fixit#1{}

\def\dd{{\rm d}}

\def\apjs{ApJ. S}
\def\cN{{\cal N}}

\def\cS{{\cal S}}

\def\apjl{ApJL}

\def\aap{A\&A}

\def\apjl{ApJL}

\usepackage{amsmath}
\usepackage{bm}
\usepackage{color}
\usepackage{graphicx}
\usepackage{hyperref}
\usepackage{enumerate}
\usepackage{soul,xcolor}
\usepackage{float}
\usepackage{cancel}
\setstcolor{red}

\begin{document}
\title{ The galaxy power spectrum from TGSS ADR1 and the effect of flux calibration systematics}
\author{Prabhakar Tiwari}
\email{ptiwari@nao.cas.cn}
\affil{ National Astronomical Observatories, CAS, Beijing 100012, China}
\author{Shamik Ghosh}
\affil{CAS Key Laboratory for Research in Galaxies and Cosmology, Department of Astronomy, \\ University of Science and Technology of China, Hefei 230026, China.}
\affil{School of Astronomy and Space Sciences, \\ University of Science and Technology of China, Hefei 230026, China}
\author{Pankaj Jain}
\affil{Department of Physics, Indian Institute of Technology, Kanpur 208016, India}
\date{\today}
\begin{abstract}
We explore the large to moderate scale anisotropy in distant radio sources using the TGSS ADR1  catalog. We use different measures, i.e. number counts, sky brightness and flux per source, for this study. In agreement with earlier results, we report a significant excess of clustering signal above the angular scale of roughly  $10$ degrees (i.e. $l\lessapprox 20-30$). We find that some survey areas have a systematically low/high flux and argue this may be the cause of the observed signal of excess power at low multipoles.
With mocks we demonstrate the effect of such large scale flux systematics and recover TGSS like excess clustering signal by assuming $20\%$ flux uncertainties over 
$\sim 10^\circ \times 10^\circ$ size patches. We argue that that TGSS at this stage, i.e. TGSS ADR1, is not suitable for large scale clustering measurements. We find that the measure, flux per source, shows evidence of isotropy for all multipoles $l > 2$ despite the presence of systematics in the data.  
\end{abstract}
 \maketitle
\section{Introduction}
\label{sc:intro}
The radio surveys are relatively wide and deep, and therefore, our best probe to explore large scale cosmological signals over intermediate redshifts $z\approx 1$ \citep{Blake:2002ac, Singal:2011,Gibelyou:2012,Rubart:2013,Tiwari:2014ni,Cobos:2013,Tiwari:2019l123,Bengaly:2019}. The NRAO VLA Sky Survey (NVSS; \citealt{Condon:1998}) and other wide surveys e.g. Sydney University Molonglo Sky Survey (SUMSS; \citealt{Mauch:2003}), TIFR GMRT Sky Survey (TGSS; \citealt{Intema:2016tgss}) etc. have been employed extensively to explore cosmology at large scale and  many interesting results have been obtained. For example the angular galaxy power spectrum from NVSS has been used to further explore $\Lambda$CDM cosmology and to extract radio galaxy biasing \citep{Blake:2004, Adi:2015nb}. It is observed that the galaxy clustering measurements from NVSS for $l>4$ fit very well with
$\Lambda$CDM with a reasonable value of galaxy bias \citep{Adi:2015nb}, this provides further justification to standard $\Lambda$CDM and evinces that the spatial distribution of radio galaxies is consistent with standard cosmological predictions. 

As radio surveys are wide and covering up to $\sim 80\%$ of the sky, therefore,  an excellent probe to large scale isotropy. Modern cosmology assumes our Universe to be statistically homogeneous and isotropic at large scales  \citep{Milne:1933CP,Milne:1935CP}. This assumption is fundamental and a basis for all our theoretical formulation. So far the observational evidence of isotropy is  best seen with cosmic microwave background (CMB) and it is uniform to roughly 1 part in $10^5$ \citep{Penzias:1965, COBE_White:1994,WMAP:2013,Planck_iso:2016}. 
The most prominent anisotropy of the CMB is a dipole signal which is of order $\sim 10^{-3}$ 
\citep{Conklin:1971,Henry:1971,Corey:1976,Smoot:1977,Kogut:1993,Hinshaw:2009}. It is commonly believed that this large dipole in CMB is caused by our local motion with respect to CMB frame \citep{Stewart:1967}. The CMB dipole signal predicts our local motion to be 369.82$\pm$ 0.11 km s$^{-1}$ in the direction, 
$l=264.021^o\pm 0.011^o$, $b=48.253\pm 0.005^o$ \citep{Planck_I:2018, Planck_III:2018}. 

The radio galaxy distribution from  NVSS and TGSS catalogs is expected to peak at redshift $z \approx 1$ \citep{Wilman:2008}, corresponding to   a comoving distance of  $\sim 3.3$ Gpc for  the $\Lambda$CDM cosmological background \citep{Planck_results:2018}. Assuming $\Lambda$CDM background power spectrum, the galaxy clustering at these distances should contribute very little to dipole, however, due to local motion, the Doppler and aberration effects are expected to produce a dipole in radio galaxies distribution \citep{Ellis:1984}. 
The measurement of this dipole signal in radio galaxies is expected to be an independent measure of our local motion after CMB, and therefore has been of great interest in scientific community.

There have been several attempts to measure the dipole from NVSS radio catalog  \citep{Baleisis:1998,Blake:2002,Singal:2011,Gibelyou:2012,Rubart:2013,Tiwari:2014ni,Tiwari:2015np}. 
Surprisingly, the velocity of our local motion as estimated from the radio dipole exceeds the velocity as calculated from the CMB dipole \citep{Singal:2011,Gibelyou:2012,Rubart:2013, Tiwari:2014ni,Tiwari:2015np}.
This excess radio dipole from NVSS is puzzling and seemingly implies that the radio galaxy distribution can be a glimpse of violation of isotropy at large scales. 
This observation of dipole  with NVSS is at least 2.3$\sigma$ \citep{Tiwari:2016adi} away from  CMB  predicted velocity  dipole and apparently a violation of isotropy and hence this particular observation has achieved significant attention. There are also other attempts to measure dipole signal in radio galaxy distribution with complementary radio surveys.  In particular \cite{Colin:2017} combined NVSS and SUMSS to achieve full sky coverage and report a dipole signal which remains roughly the same in magnitude and direction as with NVSS catalog exclusively.

It is also interesting to use other catalogues, such as the TGSS to explore large scale isotropy and to measure our local motion i.e. the dipole. The TGSS covers $90\%$ of the full sky, slightly larger sky coverage than NVSS, and it overlaps over all the NVSS coverage. TGSS catalog is apparently very much similar to NVSS and therefore attempts are made to perform similar analysis (\citealt{Bengaly:2018,Dolfi:2019,Rana:2019}[Erratum:\citealt{Rana_Erratum:2019}]) with TGSS to complement and resolve NVSS observations. However with TGSS the large scale ($l<20$)  clustering signal   is found to be relatively large  \citep{Bengaly:2018,Dolfi:2019}.
The dipole from TGSS is observed to be roughly 3 times in magnitude if compared with NVSS. It is so far not clear whether the excess clustering signal from TGSS is real physical or due to some systematics. Furthermore as the anomalous signal from TGSS disagrees with $\Lambda$CDM predictions, and also with radio clustering signal from NVSS, it further complicates the interpretation. 
Although it is likely that the observed signal in TGSS is due to systematics and NVSS remains more reliable, a clear understanding of TGSS systematics remains due. This paper is an attempt to study TGSS systematics and its reliability as compared with NVSS.

The outline of the paper is as following. We describe the TGSS and NVSS data in Section \ref{sc:data}. In Section \ref{sc:formulation} we formulate the theoretical angular power spectrum and provide expression for kinematic dipole. We present different measures of clustering i.e. number counts, sky brightness and flux per source in Section \ref{sc:obs}. In Section \ref{sc:mask} we discuss data mask and  present the clustering measurements with  different observables in Section \ref{sc:results}. In Section \ref{sc:systematics} we explore the flux systematics in data and demonstrate the effect of flux offsets using mocks. We conclude with discussion  in Section \ref{sc:conclusion}.

\section{Data}
\label{sc:data}
\subsection{The TGSS catalog}
TGSS\footnote{\href{http://tgssadr.strw.leidenuniv.nl/doku.php}{http://tgssadr.strw.leidenuniv.nl/doku.php}} \citep{Intema:2016tgss} is a continuum radio survey at 150 MHz carried out at Giant Metrewave Radio Telescope (GMRT)  \footnote{\href{http://www.gmrt.ncra.tifr.res.in/}{http://www.gmrt.ncra.tifr.res.in/}} \citep{Swarup:1991} over the period of 2010 to 2012. The survey covers roughly 90\% of the full sky north of declination $-53 \degr$ at frequency 150 MHz with median RMS brightness fluctuations 3.5 mJy/beam with approximate resolution 25\arcsec x 25\arcsec north of 19$\degr$ DEC and 25\arcsec x 25\arcsec/cos(DEC-19$\degr$) south of 19$\degr$ as the beam starts to elongate in N-S direction due to projection effects.
The primary beam size i.e. the field-of-view of GMRT at 150  GHz is 2.5-3 degree and the survey consists of more than 5000 
pointings, each having its 2.5-3 degree field of view. The available TGSS catalog, TGSS Alternative Data Release 1 (TGSS ADR1) was produced by \cite{Intema:2016tgss}. \cite{Intema:2016tgss} re-processed the survey data by using their fully automated data pipeline based on SPAM package \citep{Intema:2009phd,Intema:2009,Intema:2014}. 
 \cite{Intema:2016tgss} use  three flux  calibrators, 3C48, 3C147 and 3C286 for TGSS ADR1 processing. It is noted that TGSS observation sessions often contain pointings that are clustered together, this creates flux offsets over areas that are larger than single pointings (3-4 degrees) and can be  as large as 5-20 degrees depending on the observation strategy for particular sessions.

TGSS ADR1\footnote{throughout the paper we have used this catalog only, we loosely call TGSS ADR1 as TGSS catalog} contains 623,604 objects in total, however, it is only expected to be complete above 100 mJy ( integrated flux density\footnote{throughout the
paper we have used integrated flux density}) and there are only 307,787 sources above this flux limit. Cosmological clustering analysis can only be performed above the survey completeness and therefore we constrain our analysis with sources which have flux greater than 100 mJy.

\subsection{The NVSS catalog}
We use NVSS \citep{Condon:1998} catalog as reference, our ``good catalog". The NVSS was carried out by Very Large Array (VLA) observatory and contains $\sim$1.8 million sources with flux densities $S_{\rm 1.4GHz}>2.5$ mJy at 1.4 GHz.
The full width at half maximum angular resolution is 45\arcsec. 
The catalogue overlaps TGSS and covers  about 82$\%$ of the sky north of declination $-40\degr$. The catalogue is complete above 3.5 mJy at 1.4 GHz. However the NVSS suffers from significant systematic galaxy number density fluctuations across the sky at flux densities $S<15$ mJy \citep{Blake:2002}, which affects lower multipoles. Therefore, all the clustering analysis with NVSS so far are only carried out with sources brighter than 15 mJy \citep{Baleisis:1998,Blake:2002,Singal:2011,Gibelyou:2012,Rubart:2013,Tiwari:2014ni,Tiwari:2015np}. The clustering results from NVSS, except dipole,  agrees well with $\Lambda$CDM predictions \citep{Adi:2015nb,Tiwari:2016adi,Tiwari:2019l123}.

\begin{figure}
     \centering
     \includegraphics[width=0.48\textwidth]{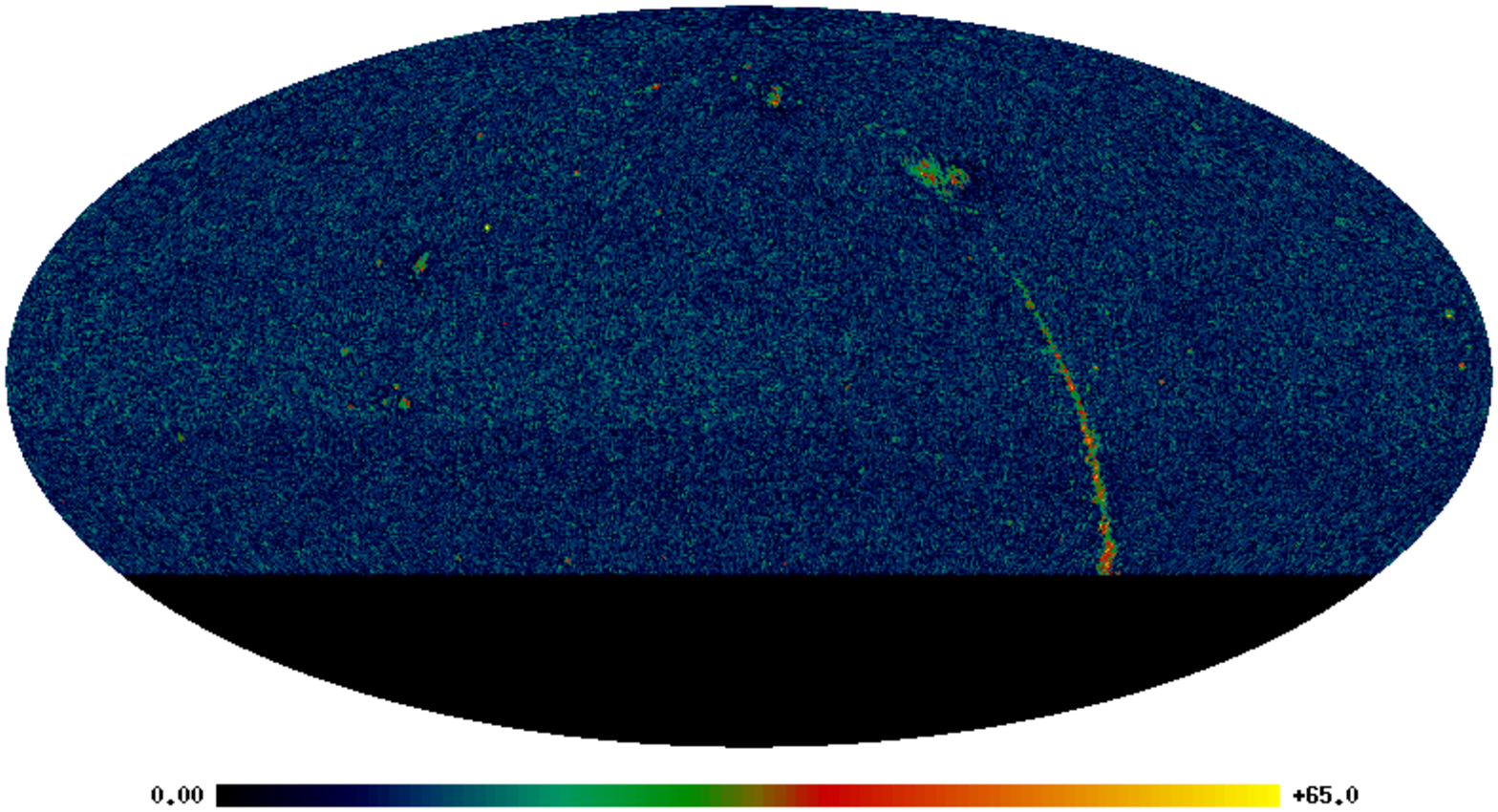}\\
     \includegraphics[width=0.48\textwidth]{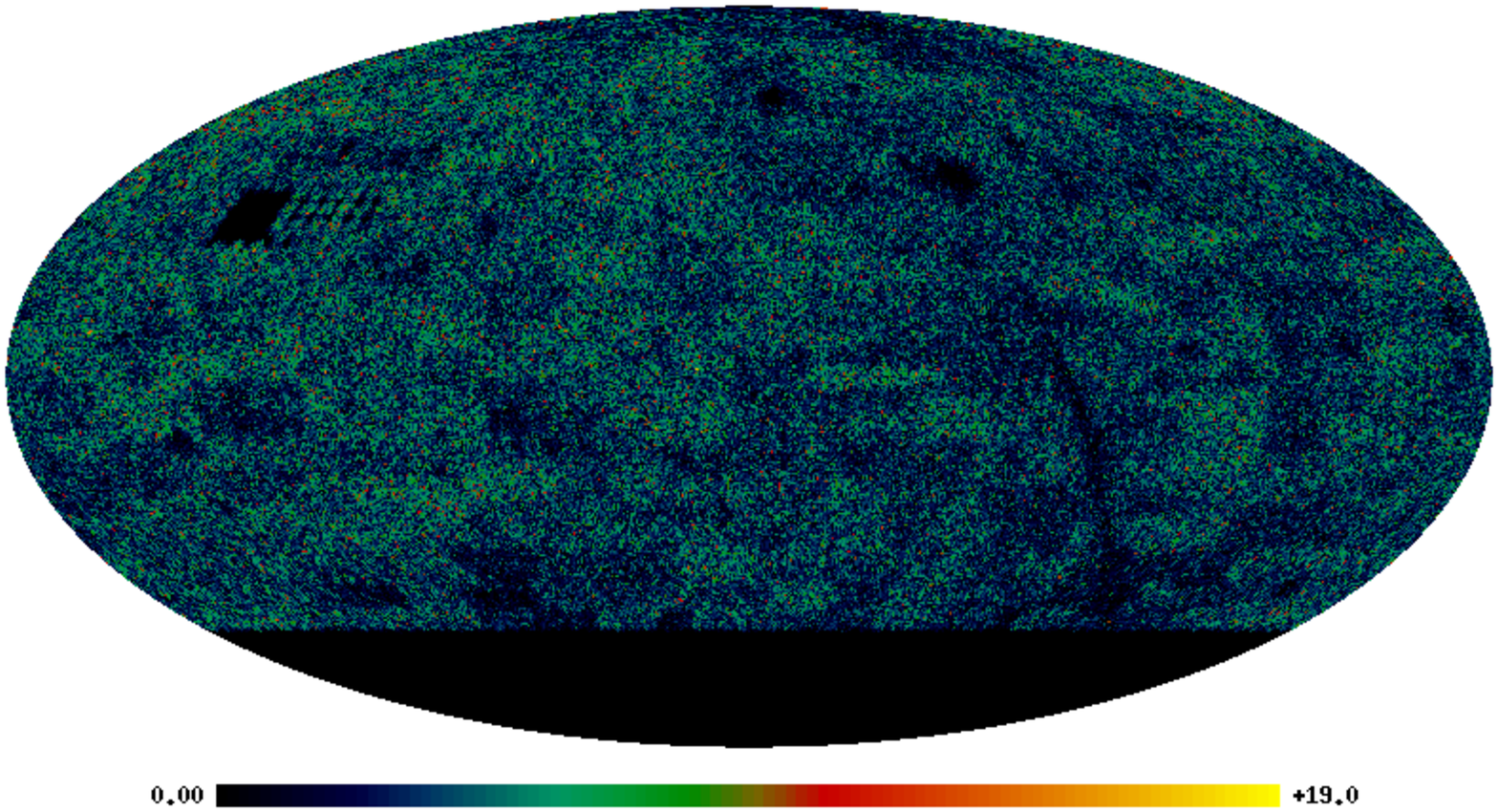}\\
     \caption{In the top panel we have plotted the NVSS catalog galaxy distribution over the sky. Bottom is the TGSS catalog galaxy distribution. The color bar for number of galaxies goes from 0 to 65 for NVSS and 0 to 19 for TGSS respectively.}
\end{figure}

\section{Theoretical formulation}
\label{sc:formulation}
\subsection{Angular power spectrum}
The clustering of galaxy spatial distribution is conventionally measured in terms of angular power spectrum, $C_l$. The observed $C_l$ can be easily connected with background matter density and thus the background cosmological model. The theoretical formulation of $C_l$ following $\Lambda$CDM scenario is as follows. Let $\cN(\hvn) =\bar \cN(1+\Delta (\hvn))$, be  the projected number density per steradian in  the direction $\hvn$. Here $\bar \cN$  is the mean number density, and  $\Delta (\hvn)$ is the projected number density contrast. Now $\Delta (\hvn)$ can be  theoretically connected to the background matter density contrast, $\delta_m(\vr,z(r))$. Here $\vr$ stands for comoving distance $r$ in direction $\hat{r}$ and $z(r)$ is the redshift corresponding to comoving distance $r$.
Next we can write the galaxy density contrast,
\beq
\delta_g(\vr,z(r)) =\delta_m(\vr,z=0) D(z) b(z),
\eeq
where $b(z)$ is galaxy biasing, $D(z)$ is the linear growth factor and $z=z(r)$. Following these we can write the theoretical $\Delta (\hvn)$,
\begin{eqnarray}
\label{eq:delta_th}
\Delta (\hvn) &=& \int _{0}^{\infty} \delta_g(\vr, z(r)) p(r) dr  \nonumber\\
              &=&  \int _{0}^{\infty} \delta_m(\vr,z=0)  D(z) b(z) p(r) dr , 
\end{eqnarray}
where, the radial distribution function, $p(r) \dd r$ is the probability of observing galaxy between comoving distance $r$ and
$(r+ dr)$. In principle $\Delta (\hvn)$ may also have some tiny additional contributions from {\bf lensing}, redshift distortions, physical distance fluctuations and from variation of radio source luminosities and spectral indices \citep{Chen:2015}. These effects are still expected to be limited to few percent on the largest scales \citep{Dolfi:2019} and can be ignored for this present work. To achieve $C_l$ from $\Delta (\hvn)$,  we expand it in terms of spherical harmonics,
\beq
\label{eq:alm}
\Delta (\hvn) = \sum_{lm}a_{lm}Y_{lm}(\hvn).
\eeq
Alternatively we can write $\al$ as,
\begin{eqnarray}
\label{eq:alm_gal}
\al&=&\int d \Omega \Delta(\hvn)  Y_{lm}^*(\hvn)\\
\nonumber &=& \int d\Omega Y_{lm}^*(\hvn) \int_{0}^{\infty} \delta_m(\vr,z=0)  D(z) b(z) p(r) dr\;.
\end{eqnarray}
The background matter density field $\delta_m(\vr,z=0)$ can be expressed as a Fourier transform of $k$-space density field $\delta_{\vk}$, as
\begin{equation}
\delta_m(\vr,z=0) =\frac{1}{(2\pi)^3}\int d^3 k \delta_{\vk}{\rm e}^{i \vk \cdot{  \vr}}\;.
\end{equation}
Using above expression we can write $\al$ as\footnote{We have substituted ${\rm e}^{i \vk \cdot{  \vr}}=4\pi \sum_{l,m} {i}^lj_l(kr) Y^*_{lm}(\hat {\vr})Y_{lm}(\hat {\vk})$, where $j_l$ is the spherical Bessel function of first kind for integer $l$}, 
\begin{equation}
\label{eq:alm_th2}
\al=\frac{{i}^l}{2\pi^2}\int dr D(z) b(z) p(r) \int d^3 k \delta_{\vk}j_l(kr) Y^*_{lm}(\hat {\vk}) \; . 
\end{equation}

Next we can write the expression for our desired angular power spectrum, $C_l$, as, 
\begin{eqnarray}
\label{eq:clth}
C_l&=&<|\al |^2> \nonumber\\
\nonumber &=& \frac{2}{\pi }\int dk k^2 P(k) \left\vert \int_{0}^{\infty} D(z) b(z) p(r) d r  j_l(kr)\right\vert^2 \\
          &=&   \frac{2}{\pi }\int dk k^2 P(k) W^2(k) \;,
\end{eqnarray}
where $P(k)$ is $\Lambda$CDM power spectrum and our background cosmology model, $W(k)=\int_{0}^{\infty} D(z) b(z) p(r) d r  j_l(kr)$ is the $k$-space window function. \\

For galaxy surveys like TGSS, NVSS, an estimate of $C_l$ corresponding to the theoretical $C_l$ given in Equation (\ref{eq:clth}) can be written as \citep{Peebles:1980},
\begin{equation}
\label{eq:cobs}
C^{\rm obs}_l=\frac{\langle  |a^{\prime}_{lm}|^2\rangle}{J_{lm}} -\frac{1}{\bar \cN}
\end{equation}
where  $a^{\prime}_{lm} =\int_{\rm survey} d \Omega  \Delta(\hvn) Y_{lm}^*(\hvn)$  and
$J_{lm}=\int_{_{\rm survey}}|Y_{lm}|^2 \dd \Omega$, here $J_{lm}$ is the approximate
correction for the partial survey region following \cite{Peebles:1980}. Note that in reality the galaxy surveys are always over the partial sky and we almost always need to construct full sky $C_l$'s from partial sky.  There are different recipes for partial to full sky recovery \citep{Peebles:1980,Starck:2013,Fourt:2013}. We are also limited by galaxy number density and thus we have shot noise in $C_l$ measurements. The term $\frac{1}{\bar \cN}$ is
deducted to remove the shot noise, assuming it is Poissonian. Furthermore, the radio sources have extended structure 
and in a source catalog the same source may have multiple entries. This has a measurable effect 
on $C_l$ \citep{Blake:2004}. The effect can be modeled as a fixed offset to $C_l$ as $\delta C_l \approx 2 e \sigma_0$, 
where $\sigma_0$ is shot noise and $e$ is a constant. The value of $e$ for TGSS ADR1 is deduced  
to be $0.09\pm0.009$ \citep{Dolfi:2019}, and for NVSS it is $0.07\pm0.005$\citep{Blake:2004}. We subtract this fixed 
offset $\delta C_l$ to obtain $C^{\rm obs}_l$. 
The  error estimate on $C_l$ due to cosmic variance, sky coverage and shot-noise is,
\begin{equation}
\label{eq:dcl}
\Delta C_l = \sqrt{\frac{2}{(2l+1) f_{\rm sky}}} \left(C^{\rm obs}_l + \frac{1}{\bar \cN}\right)
\end{equation}
where $f_{\rm sky}$ is the fraction of sky covered by  survey. The first term in the bracket is the error due the cosmic variance of the signal itself while the second term is the contribution of the shot noise. For realistic agreement between the observed signal and theory, $C_l^{\rm obs}$ can be closely approximated with theoretical $C_l$s. If the data has systematics and calibration errors the $C^{\rm obs}_l$ in reality (equation \ref{eq:cobs}) can be very different. In such a case the systematics and calibration issues can hide a physical sky signal in the data. Therefore, in equation \ref{eq:dcl} we assume that in the cosmic variance term, $C^{\rm obs}_l$ is  equal to the theoretical value of $C_l$. 

\subsection{Kinematic dipole}
Our local motion with respect to CMB or galaxies at cosmological distance  is expected to give rise to dipole anisotropy as a leading order effect. Let us assume that the velocity of our observation frame's local motion is $\vec v$. The flux density $S$ of radio sources can be modeled as a power-law distribution with the frequency $S \propto \nu^{-\alpha}$, where $\alpha \approx 0.75$ is the radio spectral index. The differential number count $n$ along a direction $\hvn$, per unit solid angle per unit flux density is usually modelled as \citep{Tiwari:2014ni}:
\beq
    n(\hvn, S)= \frac{d^2N}{d\Omega dS} = N_0 (\hvn) x S^{-1-x}, 
    \label{eq:diffncount}
\eeq
where $N_0$ is a normalization constant and have spatial dependence due to cosmological clustering signal. However, in the calculation that follows we would suppress this intrinsic cosmological variation in $N_0$ and treat it as a constant, as is customary in the literature. Then the dipole, $D_{\rm kin}$, generated in the number count, and also {\it brightness} i.e. the \emph{flux-weighted} number count, due to the local motion of our frame of observation is \citep{Ellis:1984,Tiwari:2014ni}, 
\begin{equation}
    \vec D_{\rm kin} = \left[2+x(1+\alpha)\right]\frac{\vec v}{c}.
    \label{eq:dipole}
\end{equation}
The dipole magnitude $|\vec D|$ is related to the angular power spectrum, $C _1$, as,  
\beq
C_1 = \frac{4\pi}{9} |\vec D |^2 
\label{eq:dipoleC1}
\eeq

\section{Observable and estimators}
\label{sc:obs}
\subsection{Number count}
The radio galaxy number count above some low flux cut $S_l$, where $S_l$ is above survey completeness, is one of the main observable for the estimation of galaxy clustering. One can also consider a finite upper flux $S_h$. Then, for a small patch, $r$, like a pixel on a pixelated sky map  the number count, $N_r$, can be obtained by integrating equation (\ref{eq:diffncount}) over the flux range and the area of the pixel.  It is given as \citep{Ellis:1984}, 
\beq
    N_r = N_0 \Delta \Omega_r \frac{\left( S_h^x - S_l^x\right)}{\left(S_l S_h\right)^x}.
    \label{eq:powerlawN-S}
\eeq
where, $x\approx 1$ \citep{Ellis:1984,Tiwari:2015np} and $\Delta \Omega_r$ is the area of the pixel. The number density for the pixel is $\cN_r = N_r/\Delta \Omega_r$. Note that the galaxy number density contrast is theoretically connected with background matter density and therefore this is a tracer for background power spectrum. Every survey has limited number of sources and thus the galaxy number density contrast map has finite shot noise, $1/\bar \cN$,  where $\bar \cN$ is the mean number density and can be written as:
\beq
   \bar \cN = \frac{1}{4\pi f_{\rm sky}} \frac{\left( S_h^x - S_l^x\right)}{\left(S_l S_h\right)^x} \int_\text{survey} N_0(\hvn)d\Omega 
\eeq 

In presence of position dependent flux density calibration systematics the observed flux density $\tilde S$ is related to the actual flux density by 
\beq
\tilde S = S\left[ 1 + k(\hvn) \right]\,,
\eeq
where $k(\hvn)$ is a position dependent calibration error and for simplicity we considered  this independent of source flux density $S$. If we assume that our calibration error does not vary over the area of our patch then we can replace it with a single value for the patch/pixel, $k_r$. For such a model of the flux calibration error the differential number count can be written in terms of the observed flux $\tilde S$ as:
\beq
     n(\hvn, \tilde S)= \frac{d^2N}{d\Omega d\tilde S} = N_0 (\hvn) x \tilde S^{-1-x} \left[ 1 + k(\hvn) \right]^x, 
\eeq
Then the number count for our patch with a position dependent flux calibration systematics is given by:
\beq
N_r = N_0 \Delta \Omega_r \left(1+k_r\right)^x \frac{\left( \tilde S_h^x - \tilde S_l^x\right)}{\left(\tilde S_l \tilde S_h\right)^x}.
    \label{eq:number_systematic}
\eeq
Here the upper flux limit $\tilde S_h$ and lower flux cut $\tilde S_l$ are placed on the mis-calibrated flux densities. Comparing equation \ref{eq:powerlawN-S} with equation \ref{eq:number_systematic} we see the presence of an additional position dependence coming from the calibration error.

\subsection{Brightness map}
Integrated flux density weighted number counts or sky {\it brightness} is another observable which can be used to estimate the clustering signal. In particular it has been used to measure the NVSS dipole signal \citep{Singal:2011,Tiwari:2014ni}.  Flux weighted number count, $\cS_r$, in the patch $r$, can be obtained by integrating $n(\hvn,S) \times S$ over the flux range and angles subtended by the patch to give: 
\begin{eqnarray}
\cS_r &=& \int^{S_h}_{S_l}\int_{\Delta\Omega_r} n(\hvn,S) S dS d\Omega, \\
        &= & N_0 \Delta\Omega_r \frac{x}{1-x}\left[S_h^{1-x}-S_l^{1-x}\right]\\
        &= & N_r \Delta\Omega_r \frac{x}{1-x}\frac{\left(S_l S_h\right)^x}{\left( S_h^x - S_l^x\right)}\left[S_h^{1-x}-S_l^{1-x}\right]
    \label{eq:brightness_r}
\end{eqnarray}
Note that for $S_h>>S_l$ and $x=1$, we get
\beq
\cS_r \approx S_l \log(S_h/S_l) N_r.
\label{eq:cS1}
\eeq
This suggests that the {\it brightness} map is roughly the number count map times a direction independent constant made up of the spectral index $x$ and flux limits $S_l$ and $S_h$. The angular power spectrum measurements are therefore expected to approximately remain same with brightness map. Interestingly the kinematic dipole also remains the same with brightness map. A full derivation for the kinematic dipole is given in \cite{Tiwari:2014ni} and the final expression for dipole from {\it brightness} and number count map is exactly the same for the power law model (equation  \ref{eq:diffncount}). In general, however, the $n(S)$ relationship is expected to deviate from a pure power law behaviour. In \cite{Tiwari:2014ni} the authors also explore a generalized model which provides a better fit to data. In this case we expect a generalization of equation \ref{eq:brightness_r}. The final result for the kinematic dipole with a generalized model has been worked out in \cite{Tiwari:2014ni}.

We now consider the case of position dependent flux-calibration errors. We model the miscalibration as in the previous subsection. Then for a small patch $r$ over which we can assume the calibration error to remain constant to a value $k_r$, we find:
\begin{align}
    \cS_r &= \int^{\tilde S_h}_{\tilde S_l}\int_{\Delta\Omega_r} n(\hvn,\tilde S) \tilde S d \tilde S d\Omega, \\
    & = N_0 \Delta\Omega_r\frac{x}{1-x}\left[\tilde S_h^{1-x}- \tilde S_l^{1-x}\right] \left(1 + k_r\right)^x \label{eq:flux_systematics1}\\
    &= N_r \Delta\Omega_r \frac{x}{1-x}\frac{\left(\tilde S_l \tilde S_h\right)^x}{\left( \tilde S_h^x - \tilde S_l^x\right)}\left[\tilde S_h^{1-x}- \tilde S_l^{1-x}\right].\label{eq:flux_systematics2}
\end{align}
 where $\tilde S_l$ and $\tilde S_h$ are the lower and upper limits on the observed flux.
From equation \ref{eq:flux_systematics1} it is clear that the flux-weighted number density is going to show a position dependent variation due to the contribution of the flux calibration error that varies with position.

\subsection{Flux per unit source}
\label{sec:SoverN}
An useful quantity to work with is the flux per unit source. From equations \ref{eq:powerlawN-S} and \ref{eq:brightness_r} we can find that the flux per source $\bar \cS_r$ for a patch $r$ is given by:
\beq
\bar \cS_r = \frac{\cS_r}{N_r}= \frac{x}{1-x}\frac{\left(S_l S_h\right)^x}{\left( S_h^x - S_l^x\right)}\left[S_h^{1-x}-S_l^{1-x}\right], 
\eeq
and for the limiting case $S_h>>S_l$ and $x=1$,   $\bar \cS_r \approx S_l \log(S_h/S_l)$. We note that $\bar \cS_r$ depends only on position independent constant values and does not contain $N_0$ term which contains cosmological clustering signal. The power spectrum obtained from such a map should be consistent with shot noise. However, deviations from pure power law behaviour assumed in equation \ref{eq:diffncount} is expected to lead to an anisotropy in flux per unit source map due to kinematic dipole term \citep{Tiwari:2014ni}. There may also exist other sources of anisotropy in this observable. For example, it will show anisotropy even in the case of pure power law, equation \ref{eq:diffncount}, if the exponent $x$ depends on direction.

In the case of position dependent flux calibration error $k_r$, the flux per source $\bar \cS_r$ in a patch $r$ will be independent of $k_r$ for the case of the power law model (equation \ref{eq:diffncount}). This is because the flux calibration error contribution to $\cS_r$ and $N_r$ exactly cancels out as is evident from equations \ref{eq:number_systematic} and \ref{eq:flux_systematics1}. This implies that for a simple position dependent flux calibration model of the kind discussed here the flux per source will show a signal that is consistent with shot noise up to corrections arising due to deviation from the power law model given in equation \ref{eq:diffncount}. 


\begin{table}[t]
    \centering
    \begin{tabular}{ccccc}
        \hline
        \noalign{\vskip 0.1cm}
         Identification & RA & dec & size ($\pm$RA) & size ($\pm$dec)  \\
         \hline
         \noalign{\vskip 0.1cm}
         NGC0612 & $23.5^\circ$ & $-36.5^\circ$ & $0.5^\circ$ & $0.5^\circ$ \\
         3C48 & $24.4^\circ$ & $33.2^\circ$ & $0.5^\circ$ & $0.5^\circ$\\
         3C84 & $50.0^\circ$ & $41.5^\circ$ & $0.5^\circ$ & $0.5^\circ$ \\
         unknown & $56.6^\circ$ & $-34.4^\circ$ & $0.5^\circ$ & $0.5^\circ$ \\
         3C123 & $69.3^\circ$ & $29.7^\circ$ & $0.5^\circ$ & $0.5^\circ$ \\
         PicA & $80.0^\circ$ & $-45.8^\circ$ & $5^\circ$ & $2^\circ$ \\
         unknown & $80.7^\circ$ & $-36.5^\circ$ & $0.5^\circ$ & $0.5^\circ$ \\
         M1 & $83.6^\circ$ & $22.0^\circ$ & $0.5^\circ$ & $0.5^\circ$ \\
         3C147.1 & $85.4^\circ$ & $-1.9^\circ$ & $0.5^\circ$ & $0.5^\circ$ \\
         3C147 & $85.7^\circ$ & $49.9^\circ$ & $0.5^\circ$ & $0.5^\circ$ \\
         3C161 & $96.8^\circ$ & $-5.9^\circ$ & $0.5^\circ$ & $0.5^\circ$ \\
         unknown &  $120.0^\circ$ &      $32.0^\circ$ &  $22.5^\circ$ &  $7^\circ$ \\
         HyaA & $139.5^\circ$ & $-12.1^\circ$ & $0.5^\circ$ & $0.5^\circ$ \\
         3C270 & $184.8^\circ$ & $5.8^\circ$ & $0.5^\circ$ & $0.5^\circ$ \\
         VirA & $187.7^\circ$ & $12.4^\circ$ & $2^\circ$ & $ 2^\circ$ \\
         CenA & $201.4^\circ$ & $-43.0^\circ$ & $5^\circ$ & $5^\circ$ \\
         HerA & $252.8^\circ$ & $5.0^\circ$ & $0.5^\circ$ & $0.5^\circ$ \\
         3C355 & $261.3^\circ$ & $40.6^\circ$ & $0.5^\circ$ & $0.5^\circ$ \\
         CygA & $299.9^\circ$ & $40.7^\circ$ & $10^\circ$ & $5^\circ$ \\
         CasA & $350.9^\circ$ & $58.8^\circ$ & $5^\circ$ & $5^\circ$ \\
         
         \hline
    \end{tabular}
    \caption{The 20 sites were identified as bright, extended radio sources and were masked with a rectangular patch. The center and size of the patches are listed here.}
    \label{Tab:masked}
\end{table}

\section{Data Analysis Procedure}
\subsection{Mask}
\label{sc:mask}
\begin{figure}
     \centering
     \includegraphics[width=0.32\textwidth]{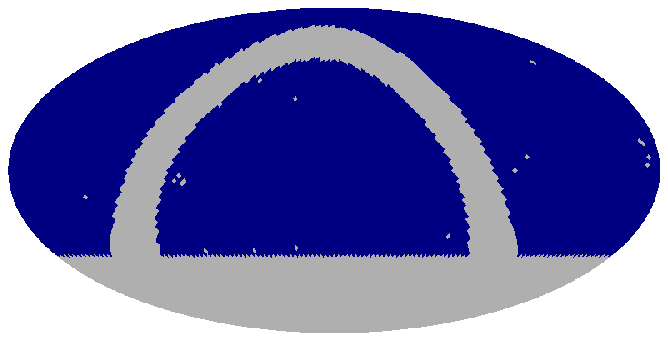}\hfill
     \includegraphics[width=0.32\textwidth]{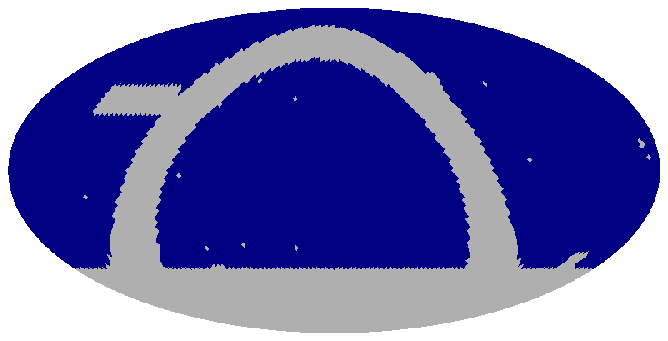}\hfill
     \includegraphics[width=0.32\textwidth]{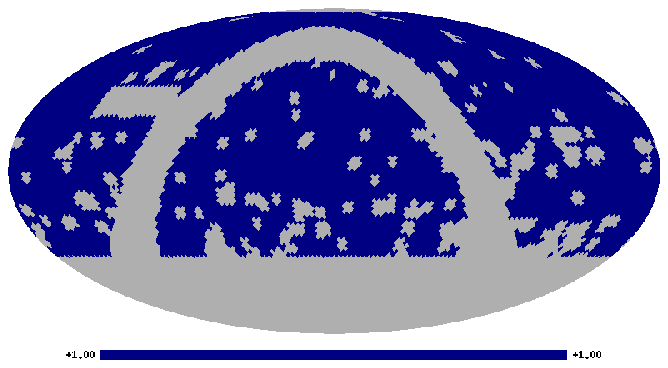}
     \caption{In the top panel we have shown the NVSS sky mask used in our analysis. Below that is the TGSS sky mask 1 with galactic plane and 20 extended sources removed and the bottom figure is the TGSS sky mask 2 which additionally removes noisy patches. Note that masks also account for missing patches in the catalog survey area.}
     \label{fig:mask}
\end{figure}

Both NVSS and TGSS catalogs have some bad data sites which we remove before performing any clustering analysis. For the NVSS catalog \cite{Blake:2002} identified 22 extended and bright sources that needed masking. We also mask the galactic plane by removing all sources with galactic coordinate latitude $|b|<10^\circ$. The resulting effective mask is shown in the top panel of  figure \ref{fig:mask}. 

For the TGSS  we mask 20 extended and bright radio objects from the catalog. The locations and sizes of these local masking is given in table \ref{Tab:masked}. The galactic plane is masked with galactic latitude $|b| < 10^\circ$. Furthermore we apply some more  stringent additional masks on TGSS. We also remove the sky with declination $\delta <-45^\circ$, in this region the RMS noise is higher than 5 mJy/beam. We show the mask thus obtained in the middle panel of figure \ref{fig:mask}. We name this mask as TGSS mask1. \cite{Intema:2016tgss} also provides the RMS noise map for TGSS. We use this to mask all regions where the noise  exceeded 5.5 mJy/beam.
We further mask $\delta <-40^\circ$ and $\delta >85^\circ$. The mask thus obtained is shown in the bottom panel of figure \ref{fig:mask}. We name this mask as TGSS mask2. 

\subsection{$C_l$ Estimation}
Throughout this work we use HEALPix \footnote{https://healpix.sourceforge.io} \citep{Gorski:2005} pixelization scheme to produce equal area pixels on spherical surface. We next populate the map with TGSS, NVSS catalogs and this gives the number count map $\cN(\hvn)$ i.e. the number of sources in a pixel in direction $\hvn$. We use an Nside $=32$ HEALPix grid to generate our number count map. Similarly we add the flux densities of all the radio sources above the flux threshold in a pixel to obtain the brightness value of that pixel. Dividing the brightness value of a pixel by the number count of the pixel we produce the flux per source maps. For the number count and brightness we obtain the mean for all unmasked pixels and then use it to calculate the contrast in number density or brightness. This removes the monopole from these maps. From this point on we will refer to the number density contrast and brightness contrast maps as number density map and brightness maps for brevity.

For the extraction of the power spectrum we use two different method for consistency. We use the \cite{Peebles:1980} method, shown in equation \ref{eq:cobs}, to obtain the reconstructed full sky $C_l$s from the different masked maps. The other method we employ is the iSAP inpainting scheme \citep{Starck:2013,Fourt:2013} to reconstruct the full sky power spectra. We have used the  default setting of the iSAP inpainting package to in-paint the missing portions of the sky. The recovery from both method is found to be roughly consistent. Furthermore, all the $C_l$s are noise de-biased by subtracting the shot noise contribution. Since the shot noise is large, for some multipoles this correction gives a negative $C_l$. Also using equation \ref{eq:dipoleC1} one can calculate the dipole amplitude.

The theoretical power spectra used for comparison is calculated using Planck cosmological parameters \citep{Planck_results:2018} following equation \ref{eq:clth} and  the galaxy bias $ {b(z) = 0.33 z^2 + 0.85 z +1.6}$, 
the radial distribution $p(r)$ i.e. $ N(z) \propto z^{0.74} \exp \left[ - \left(\frac{z}{0.71}\right)^{1.1} \right]$
detailed in \citet{Tiwari:2016adi}  and  an approximate parametric form for growth factor 
$D(z)=0.0005 z^4-0.0134 z^3 + 0.1185 z^2 - 0.4800 z + 1.0$. We use equation \ref{eq:dcl} to obtain the error for our reconstructed power spectra. For each of the number density and brightness maps we use the $C_l^{\rm theory}$ in place of $C_l^{\rm obs}$ in equation \ref{eq:dcl}. We use the mean number density and the property of Poisson distribution to get the shot noise contribution for the density contrast power spectra. The brightness maps however do not have a Poissonian distribution. To obtain the shot noise contribution we need to generate random sample maps from the distribution seen in the data. For this purpose we produce 1000 random sampled maps from the distribution of the brightness values in different pixels of the data. Random sampling the original brightness maps ensure that the random maps generated have the same distribution as the original distribution found in the data. We obtain the shot noise estimate to the brightness power spectrum from these random maps. For the flux per source power spectra we expect power spectra consistent with shot noise. So it is only justified to compare it with the power spectrum of shot noise in flux per source maps. We generate 1000 random sample from the flux per source maps just like the brightness maps. For flux per source we look at the quantity $C_l/\text{shot noise}$.

\section{Observed $C_l$ and dipole}
\label{sc:results}

\begin{figure}
\centering
\includegraphics[width=0.5\textwidth]{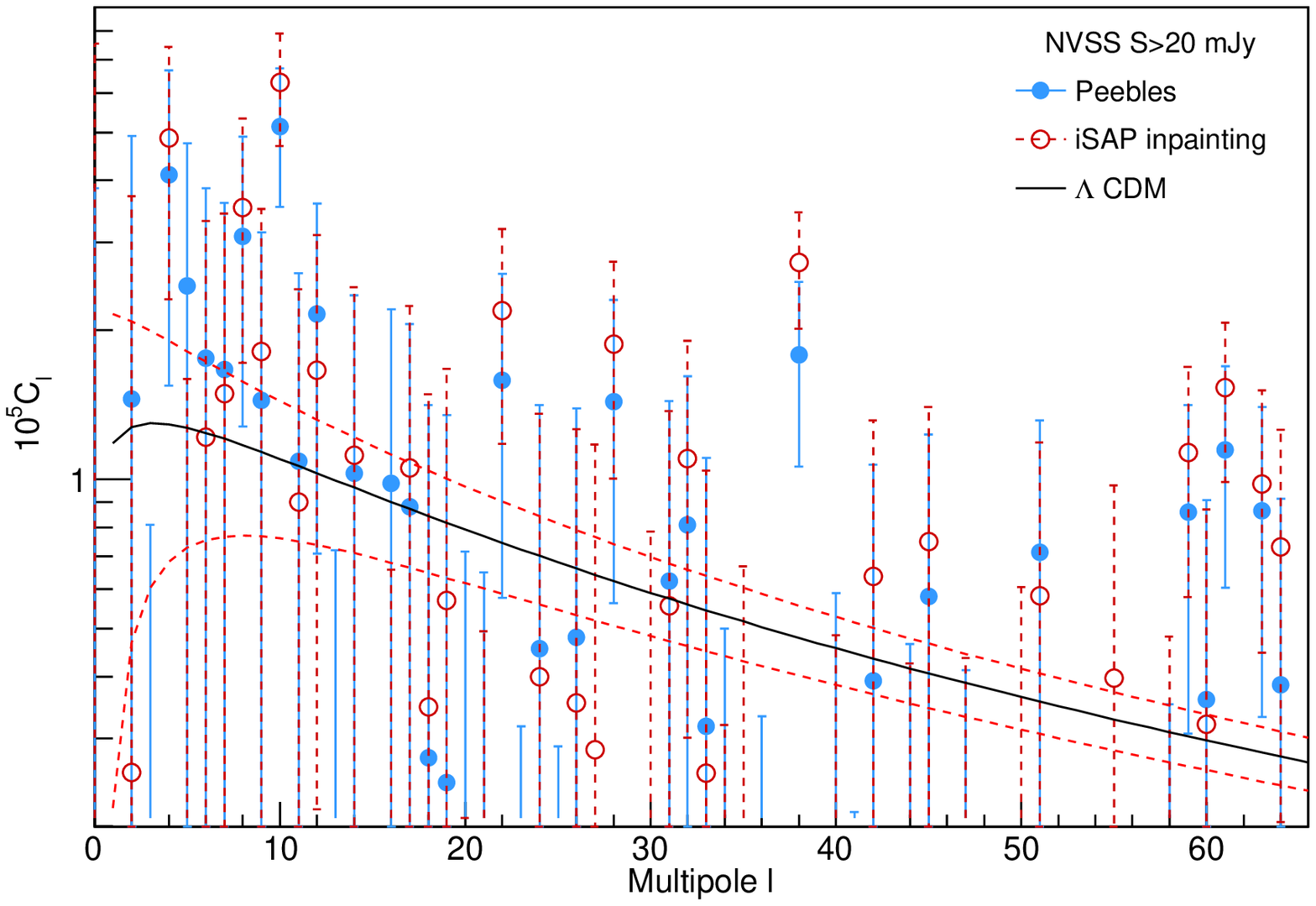}\\
\includegraphics[width=0.5\textwidth]{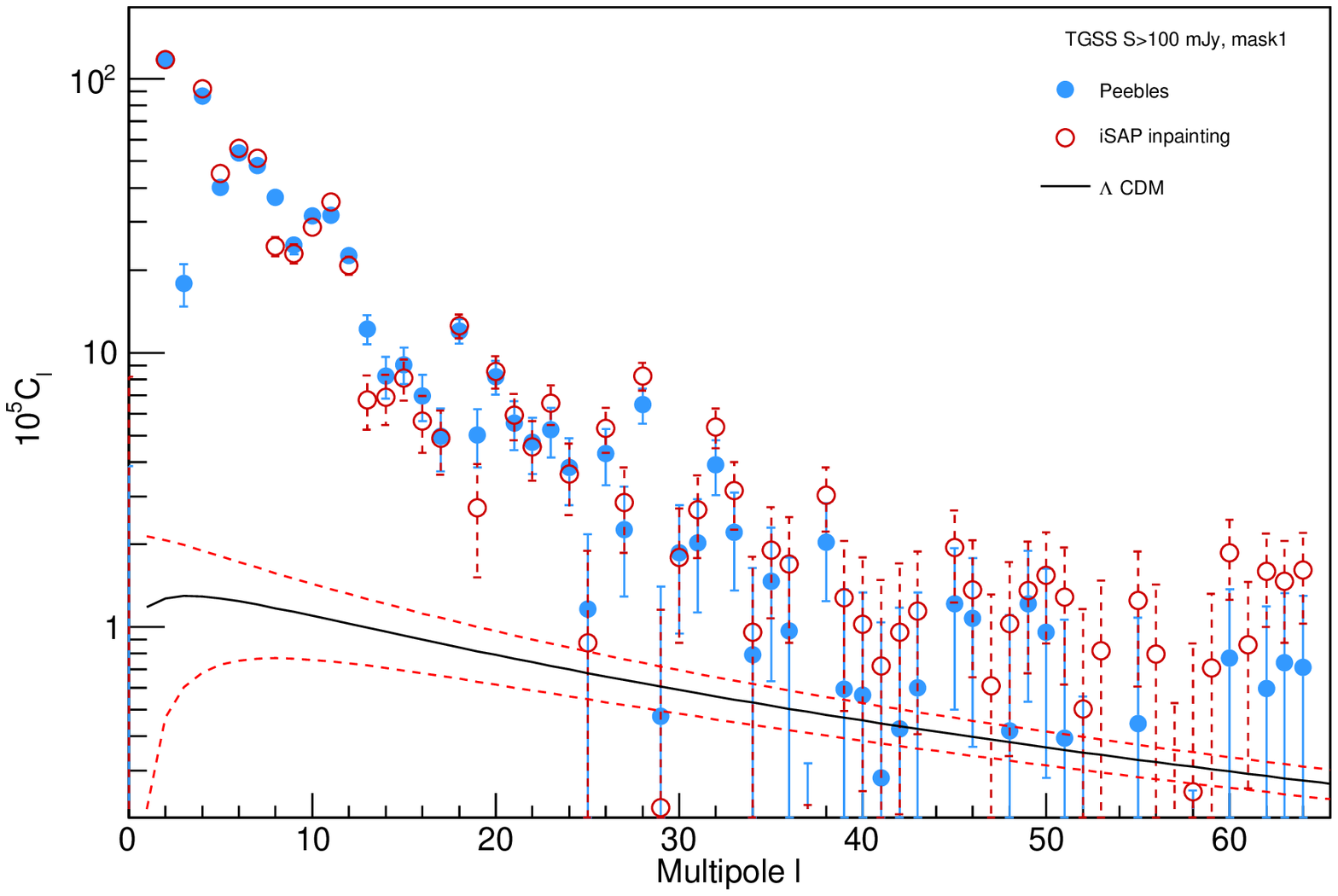}\\
\includegraphics[width=0.5\textwidth]{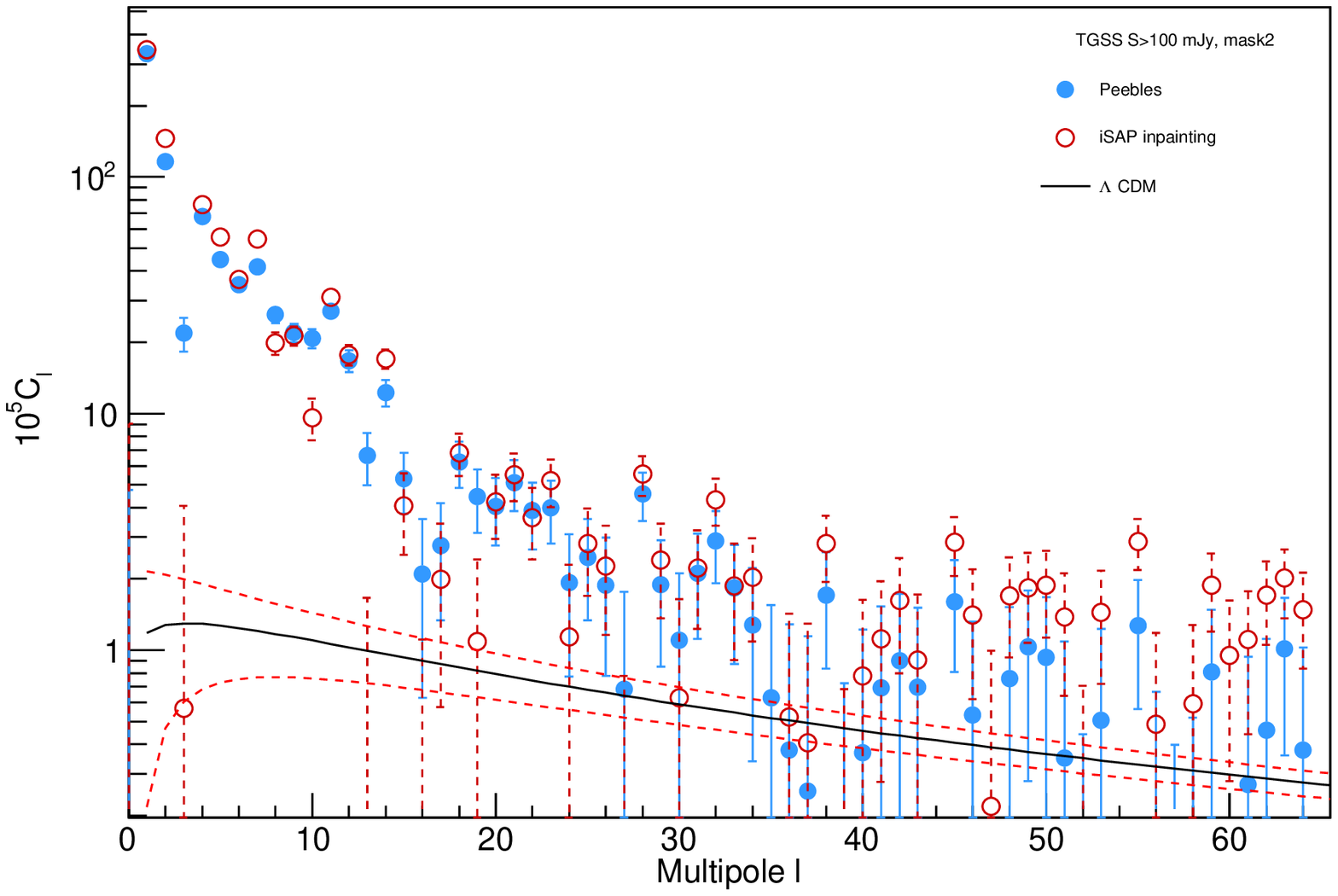}\\
 \caption{The angular power spectrum, $C^{\rm obs}_l$,   with NVSS and TGSS number density contrast map. The upper panel is NVSS ($S>20$ mJy), middle is TGSS  with mask1 and bottom is TGSS with mask2. The flux limit for TGSS map is with flux density cut $S>100$ mJy. The solid black line is theoretical prediction following galaxy bias $b(z)$, and radial distribution $p(r)$, schemes from \cite{Adi:2015nb}. The doted red lines are one sigma limits for the same. The blue (azure) and red points with error bars are $C^{\rm obs}_l$ reconstructed from partial sky following equation \ref{eq:cobs} and iSAP inpainting scheme \citep{Starck:2013,Fourt:2013}. The TGSS $C^{\rm obs}_l$ below $l\approx 20$ are significantly high.}
\label{fig:Cln}
\end{figure}

A detailed analysis with TGSS ADR1 catalog has been performed in \cite{Bengaly:2018,Dolfi:2019} who have reported their results on galaxy number density angular power spectrum, $C_l$ and dipole.  Both these studies report  $C_l$ to be significantly high below $l\lesssim30$. On the contrary the $C_l$'s from NVSS as computed by \cite{Blake:2002ac,Adi:2015nb,Bengaly:2018,Dolfi:2019} are found to be roughly consistent with $\Lambda$CDM for $l>1$. The dipole amplitude from NVSS  is more than 2 sigma disagreement with $\Lambda$CDM predictions \citep{Tiwari:2014ni,Tiwari:2015np,Tiwari:2016adi} whereas the TGSS ADR1 dipole according to \cite{Bengaly:2018}, is almost 5 times larger than the prediction. The \cite{Bengaly:2018} doubt this excess signal to be physical and say that their results may be due to systematics in data. As we are focused on systematics in this work, and the number density clustering for TGSS and NVSS has already been analysed by several authors we avoid to present detailed number density results, however, for consistency and completeness of this work we show $C_l$ from TGSS and NVSS in figure \ref{fig:Cln}. We notice from figure \ref{fig:Cln} that while NVSS is very much fitting with $\Lambda$CDM model, the $C_l$'s from TGSS ADR1 are significantly high below $l\approx 20$. The TGSS results we show in figure \ref{fig:Cln} are roughly consistent with \cite{Bengaly:2018} and \cite{Dolfi:2019}.

The lower multipoles up to $l\approx 20$ are high for TGSS ADR1 catalog. The dipole amplitude corresponding to $C^{\rm obs}_1$ is $|\vec{D}|= 0.05$ for $S>100$ mJy with mask1. 
The shot noise for the same flux cut is $3.7\times 10^{-5}$, this gives an error estimate on dipole magnitude euqal to $0.003$. Although this value of dipole is lower than the value $0.070\pm0.004$ obtained in \citep{Bengaly:2018}, however, it remain 3 to 5 times higher than the value obtained with NVSS for different flux cuts \citep{Tiwari:2014ni,Tiwari:2016adi}. The results with mask 2, shown in the bottom panel of figure \ref{fig:Cln}, also show very large $C_l$ values for low multipoles $l<10$. Only for $l\ge 10$ we find some smaller values of $C_l$ for some of the multipoles. Hence we find that mask 2 does not show much improvement in results and we do not pursue it further.

\subsection{$C_l$'s with Brightness map}
\label{ssc:cl_s}
\begin{figure}
\centering
\includegraphics[width=0.5\textwidth]{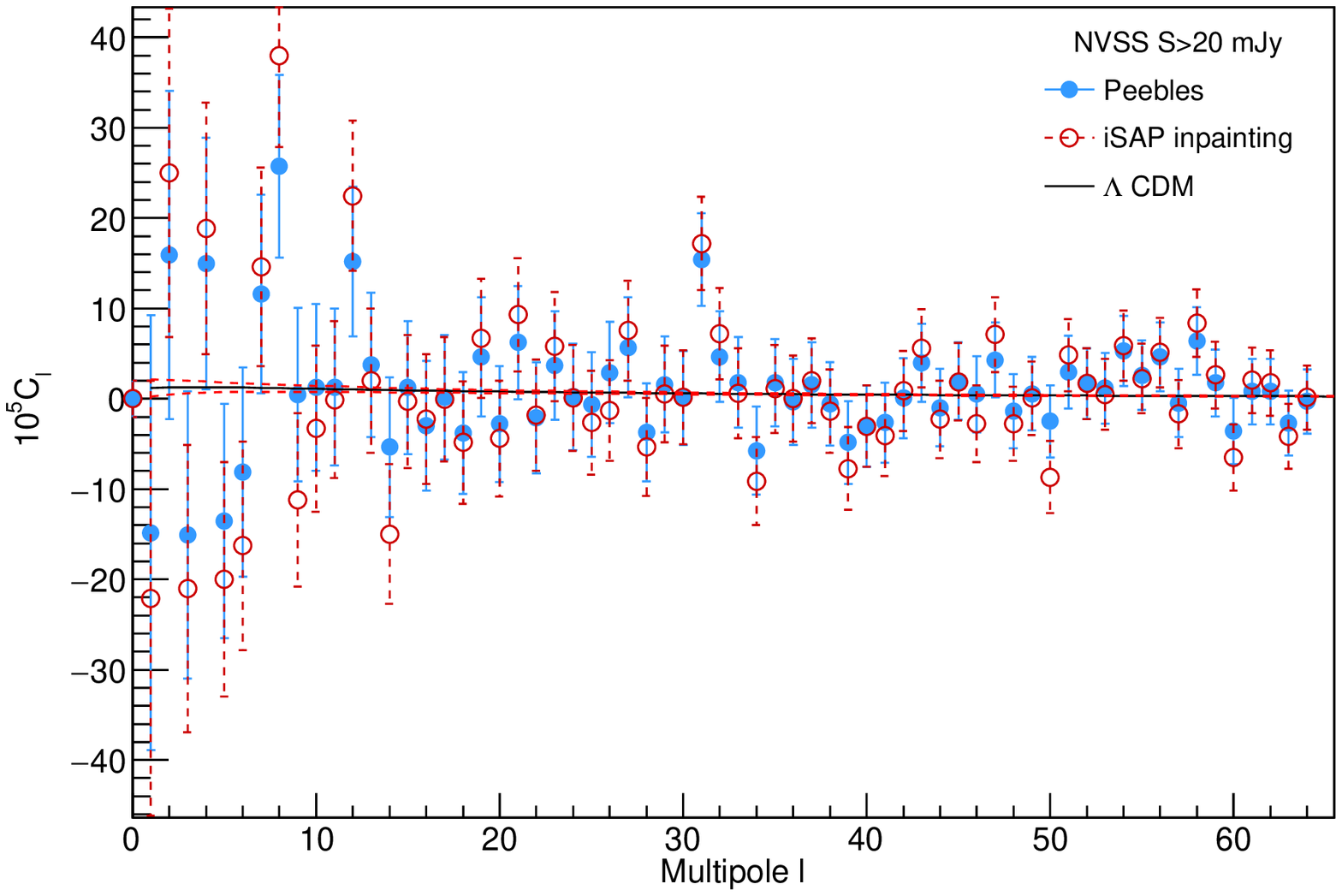}\\
\includegraphics[width=0.5\textwidth]{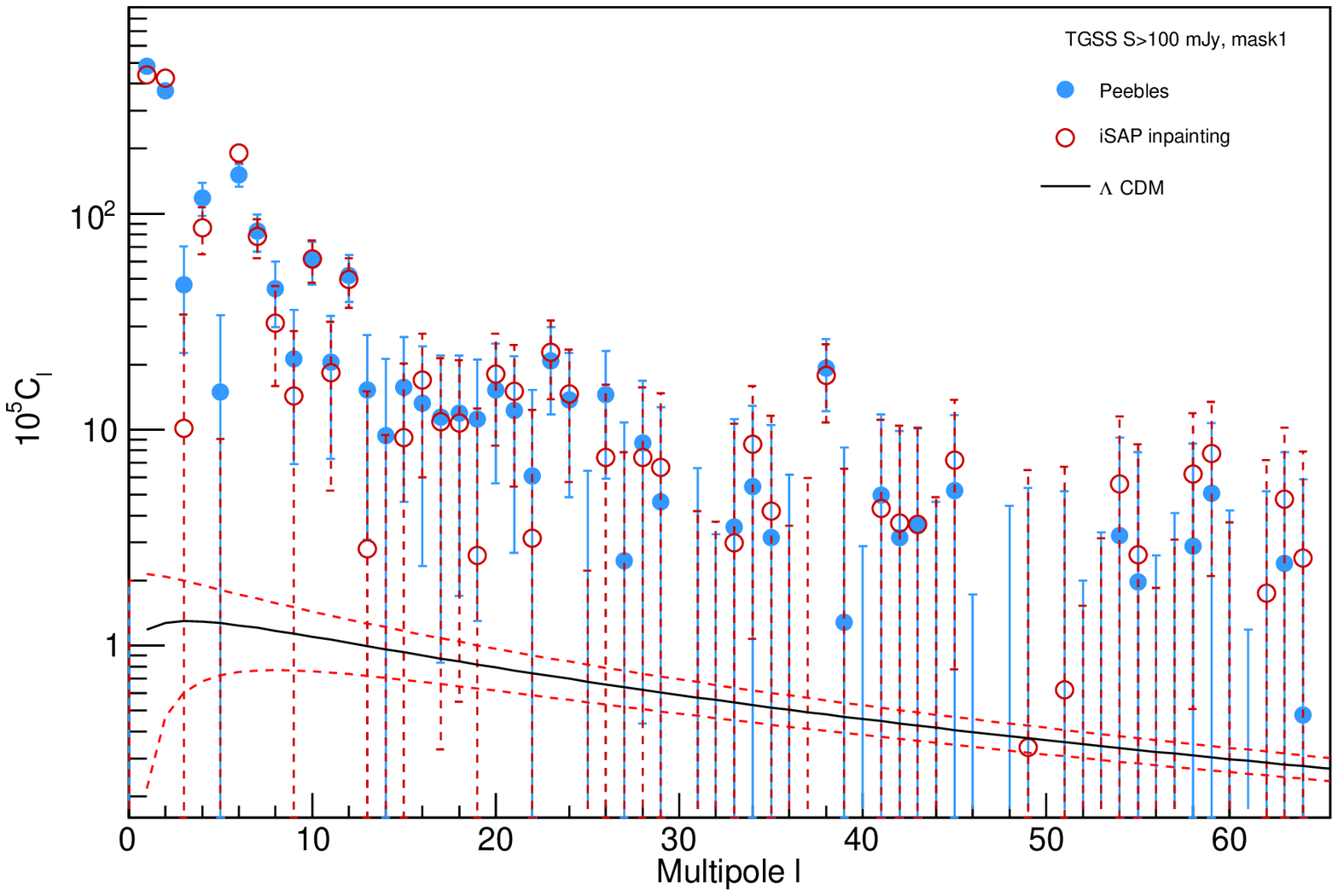}\\
 \caption{The angular power spectrum, $C^{\rm obs}_l$,   with NVSS (top) and TGSS (bottom, mask1) {\it brightness} contrast map. Again the TGSS $C^{\rm obs}_l$ are significantly high at low $l$. Other details are same as in figure \ref{fig:Cln}.}
 \label{fig:Cln_flux}
\end{figure}

The {\it brightness} map is the flux weighted number count map. This map is roughly the number density map times a constant. The results with {\it brightness} maps are shown in figure \ref{fig:Cln_flux}. The shot noise in {\it brightness} maps is much higher as compared with number density maps and the $C_l$'s are more scattered. Even so the NVSS data points are around the  theoretical predictions. However the $C_l$'s from TGSS ADR1 are systematically high. The TGSS power spectrum obtained from the brightness maps show a trend that is similar to the one seen the TGSS number density power spectrum. 

\subsection{$C_l$'s with flux per source}

\begin{figure}[h]
\centering
\includegraphics[width=0.5\textwidth]{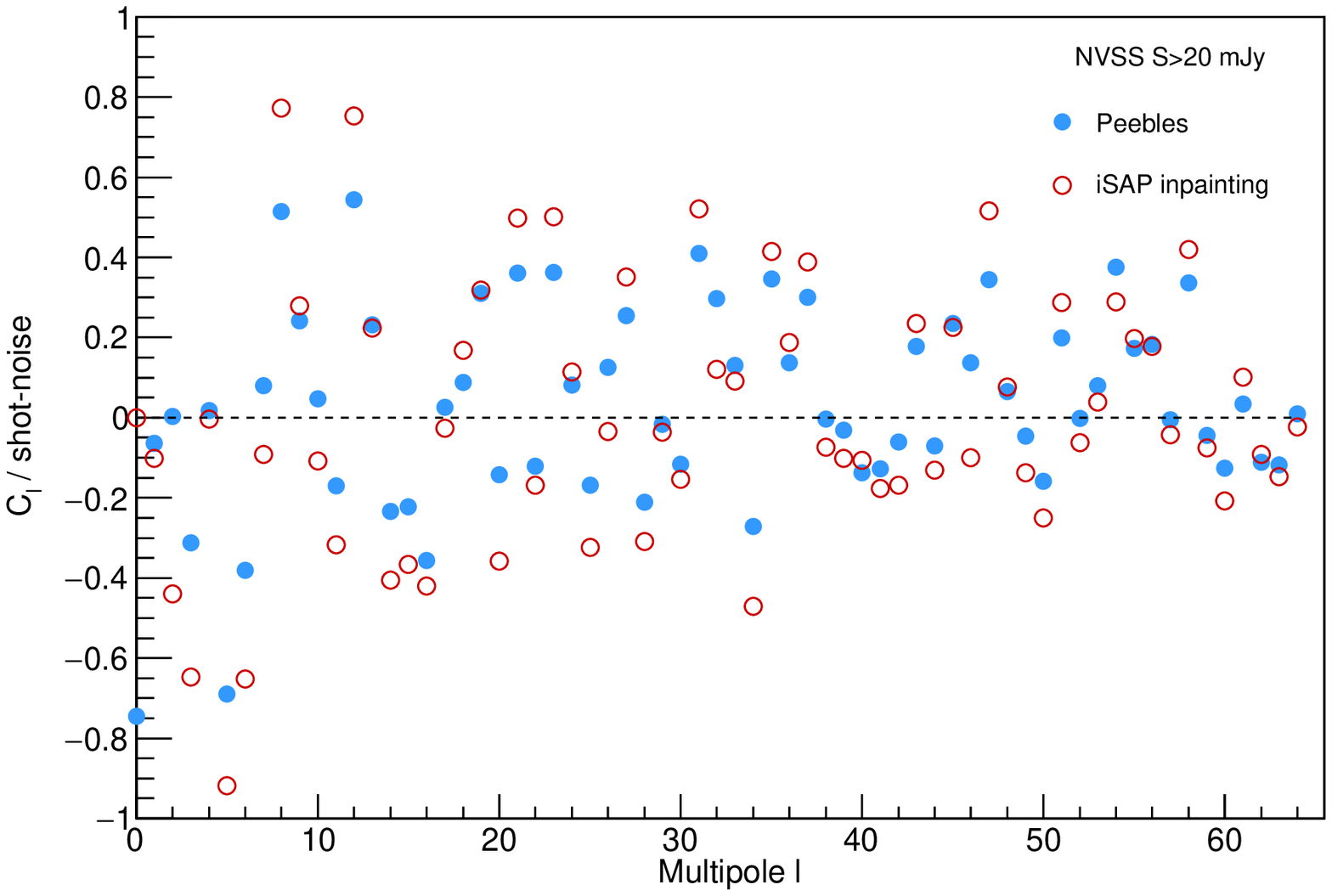}\\
\includegraphics[width=0.5\textwidth]{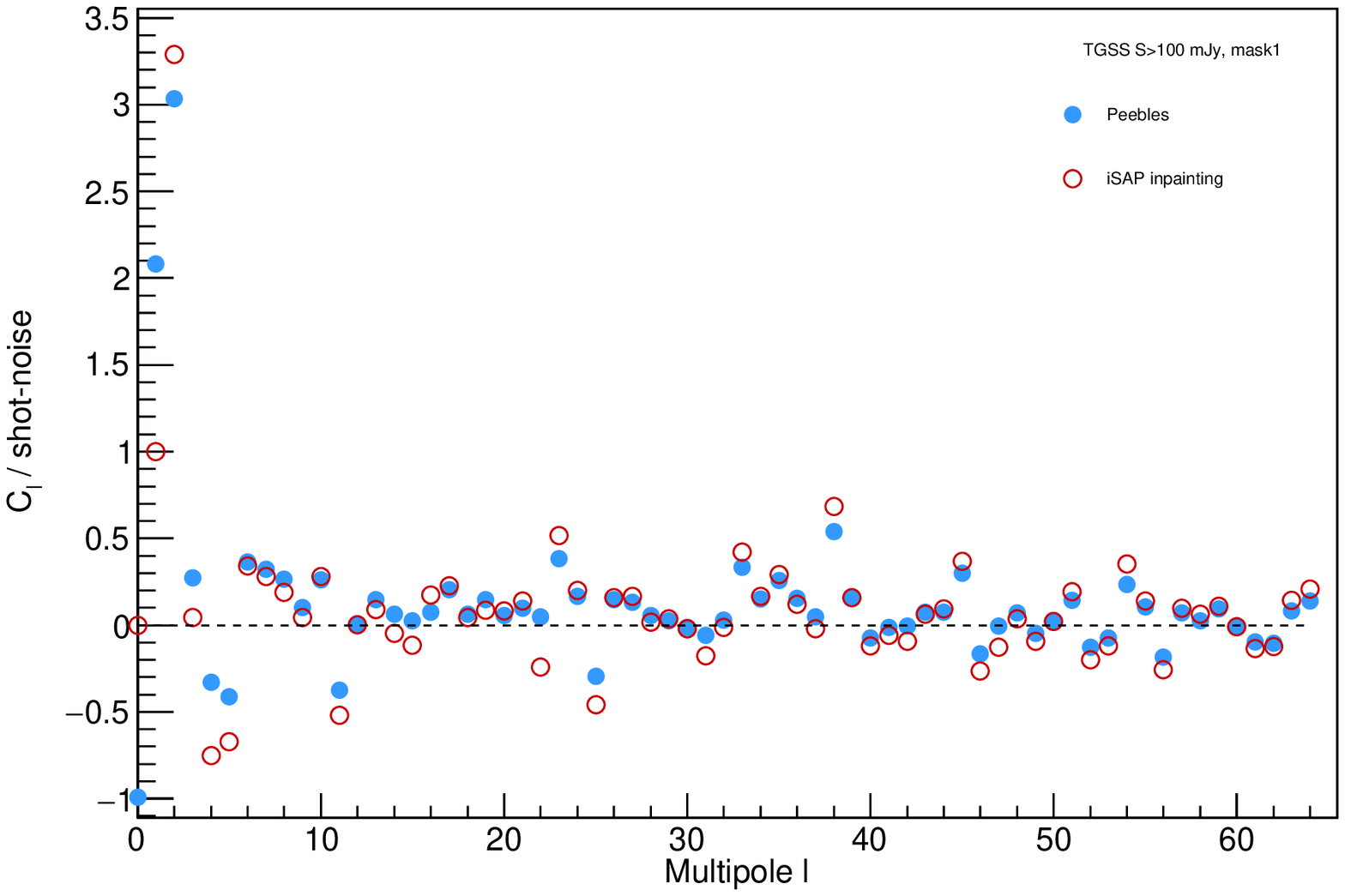}\\
	\caption{The angular power spectrum,  $C^{\rm obs}_l/$ shot-noise  with NVSS and TGSS flux per source contrast map. The figure at top is obtained using NVSS catalog ($S>20$ mJy), bottom is the TGSS ADR1 ($S>100$ mJy) figure with mask1. The blue (azure) and red points are  partial to full sky reconstruction following equation \ref{eq:cobs} and iSAP inpainting scheme \citep{Starck:2013,Fourt:2013}. Note that for NVSS the observed $C_l$'s for all values of $l$ are around zero-line and very much within one unit of shot-noise. }
 \label{fig:Cln_SpN}
\end{figure}

Our main aim in this work is to understand the TGSS ADR1 observed high clustering signal at low $l$ and to check if these are physical or due to some systematics. With figures \ref{fig:Cln} and \ref{fig:Cln_flux} we clearly see that the $C_l$s from the TGSS  is relatively high. As discussed in Section \ref{sec:SoverN} the observable flux per source ($S/N$) is expected to approximately consistent with isotropy.  The anisotropy is expected to arise due to the deviation of the $n(\hvn,S)$ from a pure power law behaviour, such as assumed in equation \ref{eq:diffncount}.  Alternatively the exponent $x$ in this equation may have spatial dependence which can also contribute to anisotropy.  As long as we have $x$ isotropic and a power law provides a good fit to data, the flux per source is expected to be a direction independent observable and we expect this map to show no clustering i.e. $C^{\rm obs}_l = 0 \pm$ shot noise. The results with NVSS and TGSS flux per source maps are shown in figure \ref{fig:Cln_SpN}. 

We clearly see that the NVSS map is showing clustering signal close to zero as expected. Remarkably even the TGSS ADR1 map shows results consistent with isotropy for all $l$'s excluding $l=1,2$. Even the $l=1,2$ do not show very large values with our mask1. The largest deviation is seen for $l=2$ which deviates by approximately $ 3$ sigmas . For higher multipoles, $l>2$,  this observable shows results nearly consistent with isotropy. This implies that to a good approximation the power law assumed in equation \ref{eq:diffncount} provides a good description of data with direction independent value of $x$. Hence despite the presence of a very significant signal of anisotropy we find that this observable shows approximate consistency with isotropy.

\section{Flux systematics}
\label{sc:systematics}

\begin{figure}
    \centering
    \includegraphics[width=0.5\textwidth]{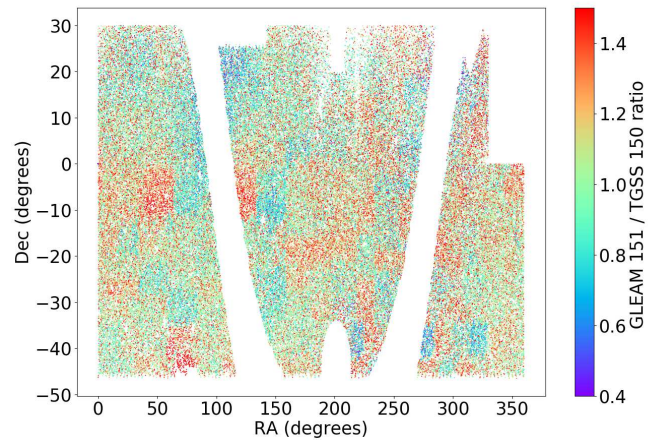}\\
    \caption{GLEAM/TGSS flux distribution with TGSS flux $>100$ mJy. Values lower than 0.4, and higher than 1.5 are drawn as 0.4 and 1.5, respectively. We notice that the flux ratio is systematically high and low in some survey regions.}
    \label{fig:gleam2tgss}
\end{figure}

\begin{figure}
\centering
\includegraphics[width=0.5\textwidth]{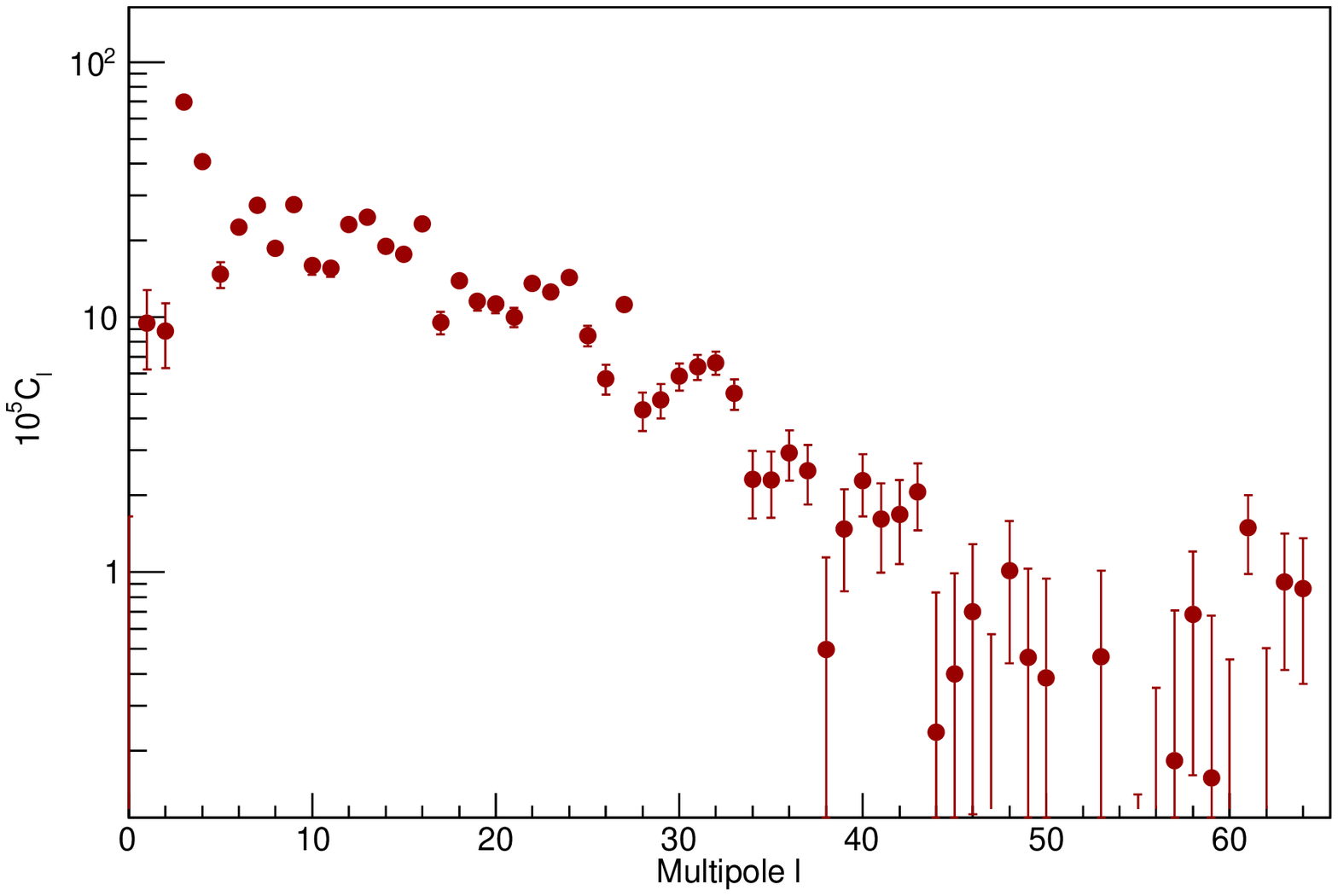}\\
\includegraphics[width=0.5\textwidth]{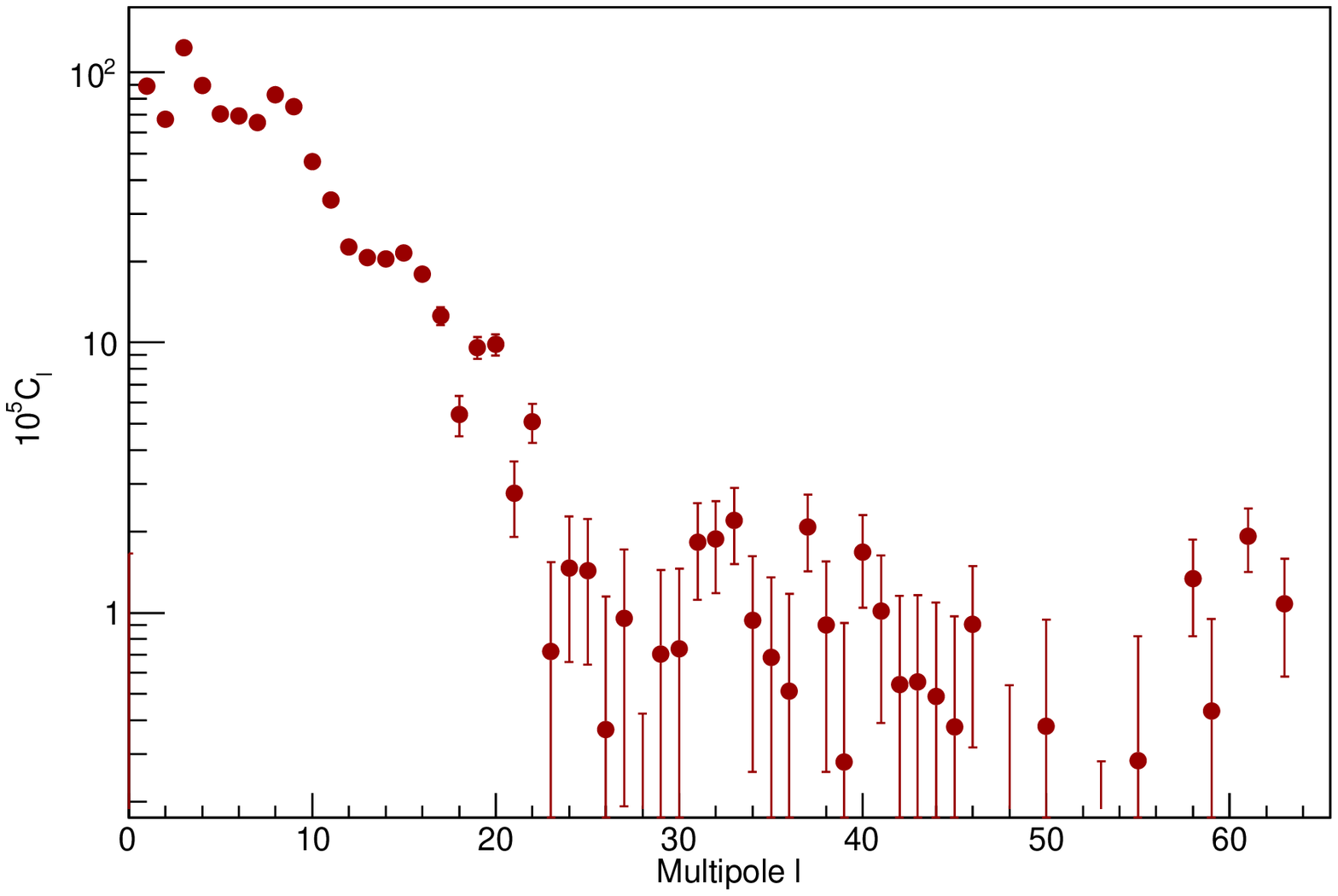}\\
 \caption{The number density contrast map  $C^{\rm obs}_l$'s with mock catalog after adding 20 percent flux offsets over a scale of $\approx 7^\circ$ (top) and $\approx 14^\circ$ (bottom)  following Gaussian distribution.}
 \label{fig:Cln_mock}
\end{figure}

\begin{figure}
\centering
\includegraphics[width=0.5\textwidth]{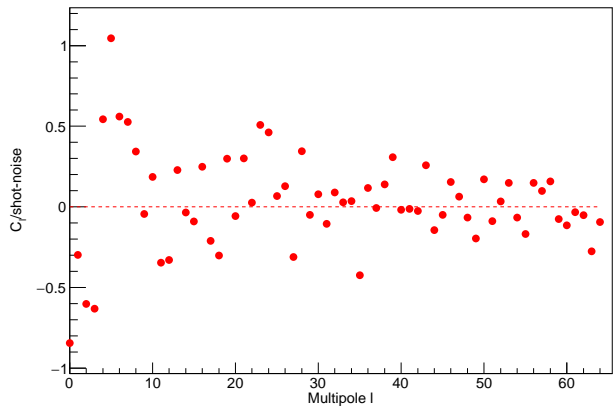}\\
\caption{The flux per source contrast map  $C^{\rm obs}_l/${\bf shot-noise} with mock catalog after adding 20 percent flux offsets over a scale of $\approx 7^\circ$.}
 \label{fig:ClSpN_mock}
\end{figure}

The TGSS data page \footnote{\label{dpage}\href{http://tgssadr.strw.leidenuniv.nl/doku.php?id=knownproblems}{http://tgssadr.strw.leidenuniv.nl/doku.php?id=knownproblems}} lists the known issues and problems with ADR1 catalog. Here the TGSS team particularly mentions that ``some areas have a systematically low flux, sometimes even 40-50 percent". 
The TGSS team\textsuperscript{\ref{dpage}} has demonstrated the flux systematics by comparing the fluxes from TGSS with GaLactic and Extragalactic All-sky Murchison Widefield Array (GLEAM) survey \citep{Hurley:2017gleam}. We plot GLEAM/TGSS  ratio in figure \ref{fig:gleam2tgss}  with mask and flux cuts used in this work. We notice that indeed the GLEAM/TGSS flux ratio is systematically high and low over the  sky. There are several 5-10$^\circ$ size regions where the TGSS flux seems to be significantly high or low. It is important to remember that the GLEAM catalog flux values are not an absolute standard and may have errors of their own.
 We learn from Huib Intema (private communication) that  in most cases, large flux offsets between TGSS and GLEAM can be traced back to particular bad observing sessions in TGSS. These large scale flux offsets could be a potential reason for large clustering signal at low multipoles. 
To demonstrate the effect of large scale flux calibration issues we mock the TGSS ADR1 catalog and then introduce random flux uncertainties. We do this as follows. First we generate RA, Dec positions randomly on sky and then to each source assign 
flux density following $n(S)\propto S^{-2}$. 
As a result we obtain a catalog with no clustering signal but with a constant shot noise signal. 
We tune the number density such that for flux cut $S>100$ mJy the catalog mimics the number density of 
TGSS ADR1. Next we introduce the large scale flux systematics. As discussed earlier TGSS observation sessions contain pointings that are clustered together. Depending on observation strategy, this 
may  creates flux offsets on scales 5-20 degrees. To mimic flux offsets on several degree scale we conventionally 
use HEALPix with NSIDE $=8$.  The NSIDE $=8$ corresponds to rectangular grids (pixels) $\approx 7^\circ \times 7^\circ$ 
over the sky. 
For each grid we next generate a random number following a Gaussian distribution with a mean of zero and 
standard deviation of $0.2$. We update the fluxes in mock catalog adding grid value times 
flux i.e. $S_{\rm new} = (1 + {\rm Gaus} (0,0.2))\times S$.  This is to introduce large 
scale (i.e. $7^\circ$ scale ) flux calibration offsets. Next we apply the 100 mJy flux cut and calculate $C_l$. 
Note that in our mocks we have generated the fluxes with lower limit of 10 mJy, this is to ensure that we do not miss any source 
after introducing flux offsets and then applying 100 mJy flux cut.
We have shown the results thus obtained in figure \ref{fig:Cln_mock}. We notice that if we add 20 percent random flux noise over a scale of $7^\circ$ then we can have excess $C_l$'s very similar to those  observed with TGSS ADR1. We repeat the same analysis with $NSIDE=4$ to inspect the effect of larger size flux offsets. We notice that large size flux offsets contribute to low multipoles more efficiently as expected. We use same mocks, with flux offsets, and obtain the flux per source power spectrum. The results are shown in figure \ref{fig:ClSpN_mock}. We notice that the effect of flux offsets remains insignificant for flux per source map and the observed angular power spectrum remains very much isotropic. This is consistent with the observation in figure \ref{fig:Cln_SpN}. The flux offsets proposed in this section can reproduce TGSS ADR1 like angular power spectrum for all our observables.

\section{Discussion and Conclusion}
\label{sc:conclusion}
We have further explored the anomalous clustering signal which has been observed with TGSS ADR1 \citep{Bengaly:2018,Dolfi:2019}. 
We employ the NVSS catalog as standard catalog (our ``good" catalog ) and apply same analysis pipeline. The results obtained from the  TGSS are compared with NVSS to display imparity between these two. We used flux weighted number count i.e. {\it brightness } map and flux per source map to study this clustering signal in more detail. For number density maps we obtained a excess in low-$l$ power (if compared with $\Lambda$CDM predictions) similar to the results in \cite{Bengaly:2018,Dolfi:2019}. Next, we obtain the power spectrum with {\it brightness } map. Although the {\it brightness } maps turned out to be noisy, with relatively large shot noise, we again observe  a similar excess in power at low $l$ with TGSS ADR1. The NVSS clustering results even with {\it brightness} map remains consistent with $\Lambda$CDM theoretical predictions.  

We next analyse the signal of anisotropy in flux per source. In this case we find that NVSS is isotropic at all scales and TGSS ADR1 is also isotropic at all scales above $l\ge 3$. The deviation for $l=1,2$ is also found to be relatively small, roughly 2-3 sigmas. Within $\Lambda$CDM, this observable is expected to show anisotropy due to kinematic dipole if the number density deviates from a pure power law assumed in equation \ref{eq:diffncount}. In general, it may show anisotropy due to flux systematics or due to an intrinsic anisotropy in the Universe. 

We study the GLEAM/TGSS observed fluxes and notice that the TGSS catalog seems to have systematically high/low fluxes over the sky. We explore the effects of large scale flux offsets with mocks.  We find that excess clustering signal as observed with TGSS ADR1 can be reproduced if there are residual calibration offsets as manifested by GLEAM/TGSS flux comparison. These position dependent flux offsets can potentially contaminate the large scale $C_l$'s by adding large systematic errors on these scales. With the GLEAM/TGSS comparison, we clearly see the flux offsets in data, therefore the dipole signal obtained with TGSS ADR1 number density maps \citep{Bengaly:2018} is likely to have a large contribution from the position dependent  flux systematics. Furthermore these large scale flux offsets will contribute to other large scale multipoles and is the likely reason behind the low-$l$ excess power observed with number density maps.

We have shown in this work that the position dependent flux calibration systematics can lead to large excesses in the clustering observed on scales $l < 20$ with the TGSS ADR1 catalog. Any future use of the TGSS ADR1 catalog towards studying the cosmological clustering signal on large or moderate angular scales needs to be mindful of these issues.

\section{Acknowledgments}
We thank Huib T. Intema for help with the TGSS ADR1 catalog masks and survey details. Prabhakar Tiwari acknowledges the support by the PIFI (Project No. 2019PM0007) program of the Chinese Academy of Sciences and the NSFC Grants 11720101004 and 11673025, and the National Key Basic Research and Development Program of China (No. 2018YFA0404503). The work of Pankaj Jain is supported by a grant from the Science and Engineering Research Board (SERB), Government of India.

\bibliography{master}

\end{document}